\documentclass[a4paper,twoside,10pt]{article}
\usepackage{amsthm}
\usepackage{amssymb}
\usepackage{graphicx}
\usepackage{cite}
\usepackage{amsmath}
\usepackage{color}

\begin{document}
\setlength{\baselineskip}{15pt}%


\centerline{\large{\bf  Detecting causality in multivariate time series via non-uniform embedding }}


\bigskip

\centerline{{Ziyu Jia$^{1}$, Youfang Lin$^{1}$, Zehui Jiao$^{1}$, Yan Ma$^{2}$, Jing Wang$^{*,1}$}}

\smallskip
\centerline{\it \footnotesize 1 School of Computer and Information Technology, Beijing Jiaotong University, Beijing 100044, PR China }
\centerline{\it \footnotesize 2 Division of Interdisciplinary Medicine and Biotechnology, Department of Medicine,}
\centerline{\it \footnotesize Beth Israel Deaconess Medical Center/ Harvard Medical School, Boston, MA, USA}

\bigskip
\bigskip
\begin{abstract}
\noindent
Causal analysis based on non-uniform embedding schemes is an important way to detect the underlying interactions between dynamic systems. However, there are still some obstacles to estimating high-dimensional conditional mutual information and forming optimal mixed embedding vector in traditional non-uniform embedding schemes. In this study, we present a new non-uniform embedding method framed in information theory to detect causality for multivariate time series, named LM-PMIME, which integrates the low-dimensional approximation of conditional mutual information and the mixed search strategy for the construction of the mixed embedding vector. We apply the proposed method to simulations of linear stochastic, nonlinear stochastic, and chaotic systems, demonstrating its superiority over partial conditional mutual information from mixed embedding (PMIME) method. Moreover, the proposed method works well for multivariate time series with weak coupling strengths, especially for chaotic systems. In the actual application, we show its applicability to epilepsy multichannel electrocorticographic recordings.

\end{abstract}

\bigskip\medskip
\noindent {\bf 1. Introduction}
\smallskip

\noindent

In recent years, various time series analysis methods have been proposed to identify interactions between complex systems\cite{refCNSNS-1}. The study of causality, in particular, has attracted wide attention of researchers. There are two classic methods in the time series causal analysis: Granger causality \cite{ref1-1} and transfer entropy \cite{ref1-2,ref1-3}. Both methods are based on time series prediction for causal analysis. In addition, the relationship between Granger causality and transfer entropy is demonstrated \cite{ref1-4}: the two methods are equivalent under Gaussian assumptions. Furthermore, Hlavackova-Schindler \cite{ref1-5} extends the equivalence of the two causality methods for generalized Gassian processes which satisfy some additional condition on probability density distributions.

With the development of multivariate state space reconstruction, different embedding schemes \cite{ref1-6,ref1-7,ref1-8,ref-add-3,refCNSNS-2} are used in Granger causality and transfer entropy. The common idea of those embedding schemes is to reconstruct the past of the whole system represented by all variables with reference to the target variable, in order to form a mixed embedding vector containing the most significant past variables to explain the target variable. Non-uniform embedding schemes are proposed to overcome the problems of arbitrariness and redundancy in uniform embedding schemes \cite{ref1-8,refCNSNS-3}. Vlachos et al. propose a causality measure based on the mixed embedding scheme for bivariate time series: the conditional mutual information from mixed embedding (MIME) \cite{ref1-9}. Kugiumtzis et al. extend the measure MIME to multivariate time series and form the partial MIME (PMIME) \cite{ref1-10}. The PMIME addresses successfully the problem of detecting direct causal effects in the multivariate variables. In addition, it is gradually applied to complex systems such as physiology \cite{ref1-11,ref1-12} and finance \cite{ref1-13,ref1-14}.

Although the causal analysis based on non-uniform embedding schemes has practical advantages, there are still some key shortcomings that need to be overcome. One shortcoming is the curse of dimensionality, which makes the estimation of mutual information inaccurate as the dimension of the embedded space increases \cite{ref1-15,ref1-16,ref1-17,ref1-18}. Another shortcoming is related to the mixed embedding vector. The greedy strategy uses a sequential forward approach to select the lagged variables and finally form the mixed embedding vector \cite{ref1-8,ref1-9,ref1-10}. That is to say, the lagged variables that have been embedded will not be changed in the mixed embedding vector. As the iteration increases, more lagged variables are selected until the final mixed embedding vector is formed. Therefore, the inaccuracy of the initial embedding will have a large impact on the results. The above shortcomings will be highlighted when there are multivariate time series of weak causal coupling strengths in practical applications.

In this paper, we put forward a new non-uniform embedding method named LM-PMIME for multivariate time series according to the low-dimensional approximation of conditional mutual information and the mixed search strategy. The main idea of the proposed method is to reduce the dimension of the embedded space by replacing the original estimate with a low-dimensional approximation of conditional mutual information. In addition, a mixed strategy, which has taken the place of the greedy strategy, was adopted as an embedded strategy to optimize the initial embedding. The proposed method works well for multivariate time series with weak coupling strengths.

The rest of the paper is organized as follows. In Section 2, we present the whole structure of multivariate non-uniform embedding in accordance with the low dimensional approximation of CMI and a mixed search strategy. In Section 3, we perform a large number of simulation experiments in order to verify the effectiveness of the proposed method. In Section 4, by analyzing the electrocorticographic (ECoG) recordings from an epileptic patient, the applicability of the proposed method to actual data is shown. Finally, a summary is presented in Section 5.

\bigskip\medskip

\noindent {\bf 2. Method}

In this section, we first introduce the traditional PMIME method. Then we expound a low dimensional approximation of conditional mutual information and  a mixed search strategy. Finally, we present the LM-PMIME method for multivariable non-uniform embedding.

\noindent{\it {2.1. PMIME Method}}

Partial conditional mutual information from mixed embedding (PMIME), a generalization of conditional mutual information from mixed embedding (MIME) for bivariate time series \cite{ref1-9}, is developed by Kugiumtzis et al. \cite{ref1-10} to estimate the directional coupling in multivariate time series. Let $K$ variables $X,Y,Z_1,\ldots,Z_{K-2}$ constitute an overall dynamical system $\{x_t,y_t,z_{1,t},\ldots,z_{K-2,t}\}{_{t=1}^n}$. Suppose that the driving subsystem is $X$ and the target subsystem is $Y$. In other words, the current value of variable $Y$ is affected by the past of  variable $X$. $Z=\{Z_1,\ldots,Z_{K-2}\}$ represent the remaining subsystems. 

We estimate the causal effect of $X$ on $Y$ conditioned by $Z=\{Z_1,\ldots,Z_{K-2}\}$. It is necessary to form a set of variables representing the past of the subsystems. The lags of $X$, $Y$ and $Z$ are sought within a range given by a maximum lag for each variable, e.g., $L_x$ for $X$ and $L_y$ for $Y$. $W_t$ is defined as the set of all lagged variables at time $t$, containing the parts $x_t,x_{t-1},\ldots,x_{t-L_x}$ of $X$ and the same for $Y$ and $Z$. It is usually assumed that the maximum lag $L$ for all variables is the same ($L_x=L_y=L_z$). The larger the value of $L$, the more lagged variables are included in $W_t$. The key step of the PMIME method is to form the mixed embedding vector $\mathbf{v_t}\in W_t$  using non-uniform embedding. Greedy forward selection and a stopping criterion are applied to the process of embedding. The detailed method is described below as follows :
\begin{enumerate}
\item An empty embedding vector $\mathbf{v{_t^0}} = \emptyset$ is initialized.
\item At the first iteration $k$ = 1, the embedding vector $w{_t^1}\in W_t$ is selected most related to $\mathbf{y_t}$:\begin{equation}
w{_t^1}=\mathop{\arg\max}_{w\in W_t}I(\mathbf{y_t} ; w)
\end{equation}where $I(.)$ represents mutual information. Mutual information is estimated by the k-nearest neighbors (k-NNs) method. Then we have $\mathbf{v{_t^1}} = [w{_t^1}]$. At the same time, $w{_t^1}$ is removed from $W_t$.
\item At the iteration $k > 1$, the mixed embedding vector is augmented by the component $w{_t^k}$ of $W_t$, giving most information about $\mathbf{y_t}$  additionally to the information already contained in $\mathbf{v{_t^{k-1}}}=[w{_t^1},\ldots,w{_t^{k-1}}]$. $w{_t^k}$ will be tested by a standard through computing the maximum conditional mutual information (CMI), $w{_t^k}=\mathop{\arg\max}_{w\in W_t}I(\mathbf{y_t} ; w|\mathbf{v{_t^{k-1}}})$, i.e., at the iteration $k = 2$, $w{_t^2}=\mathop{\arg\max}_{w\in W_t}I(\mathbf{y_t} ; w|\mathbf{v{_t^{1}}})$, where the conditional mutual information is estimated by the k-NNs estimator, and the mixed embedding vector is $\mathbf{v{_t^2}} = [w{_t^1},w{_t^2}]$. By using greedy forward method, each $w{_t^k}$ will be embedded in the already embedded vector $\mathbf{v{_t^{k-1}}}$ until the process stops. The termination criterion is quantified as :
\begin{equation}
I(\mathbf{y_t} ; \mathbf{v{_t^{k-1}}})/I(\mathbf{y_t} ; \mathbf{v{_t^k}}) >  A
\end{equation}where the threshold $A < 1$ and the general value of $A$ is 0.95 or 0.97 in \cite{ref1-9,ref1-10}. That is, the additional information of $w{_t^k}$ selected at the iteration $k$ is not large enough. The embedding process will stop and we have the mixed embedding vector $\mathbf{v_t}=\mathbf{v{_t^{k-1}}}$. And any combination of the lagged variables $X,Y,Z_1,\ldots,Z_{K-2}$ may be included in $\mathbf{v_t}$.
\item To quantify the causal effect of $X$ on $Y$ conditioned by the other variables in $Z$, the index is defined as\begin{equation}
R{_{X\rightarrow Y|Z}} = \frac{I(\mathbf{y_t};\mathbf{v{_t^x}}|\mathbf{v{_t^y}},\mathbf{v{_t^z}})}{I(\mathbf{y_t} ; \mathbf{v{_t}})}
\end{equation}where $\mathbf{v{_t^x}}$ represents the component of $X$ in $\mathbf{v{_t}}$. And it is the same with $\mathbf{v{_t^y}}$ and $\mathbf{v{_t^z}}$. The causal effect of $X$ to $Y$ depends on the components of $X$ in $\mathbf{v{_t^x}}$.
\end{enumerate}

\noindent{\it {2.2. The Proposed Method }}

\noindent{\it {2.2.1. Low Dimensional Approximation of CMI }} 

As the dimension of mixed embedding vector increases, the estimation of CMI becomes less reliable. Because of an increasing volume of state space, the estimation of entropy rates progressively decrease towards zero \cite{ref2-19}. Therefore, in order to overcome the problems caused by computing high-dimensional CMI, the low-dimensional approximation of CMI is a better alternative. The low-dimensional approximation can  improve the accuracy of conditional mutual information estimation and reduce the computational cost.

The low-dimensional approximation of CMI is studied by researchers in the field of information theory based on feature selection \cite{ref2-20,ref2-21,ref2-22,ref2-23,ref2-24,ref2-25}. Brown et al. \cite{ref1-17} emphasize that lots of feature selection heuristics are all approximate iterative maximisers of the conditional likelihood, which can be interpreted in a unifying framework of conditional likelihood maximisation under certain assumptions of independence. Consequently, the methods are summarized as a parameterized general standard:\begin{equation}
I(w;\mathbf{y_t})- \beta \sum_{w_i \in  \mathbf{v_t} } I(w;w_i)+\gamma \sum_{w_i \in  \mathbf{v_t} } I(w;w_i|\mathbf{y_t})
\end{equation}where the difference between different standards depends on the parameters ($\beta$ and $\gamma$) . For example, the JMI standard \cite{ref2-21} can be obtained with  $\beta = \gamma = 1/|\mathbf{v_t}|$. $\beta$ and $\gamma$ are different in standards such as MRMR standard \cite{ref2-23}, and CIFE standard \cite{ref2-24}. Recent studies have shown that higher-order feature interactions are considered to optimize feature selection standard. Therefore, we need to consider the second-order interactions between the features compared to Eq (4), such as $ I(w;w_j|w_i)$ \cite{ref1-18}.\begin{equation}
I(w;\mathbf{y_t})- \beta \sum_{w_i \in  \mathbf{v_t} } I(w;w_i)+\gamma \sum_{w_i \in  \mathbf{v_t} } I(w;w_i|\mathbf{y_t})-\delta \sum _{w_i \in \mathbf{v_t}}\sum_{ w_j \in \mathbf{v_t};i \not=j} I(w;w_j|w_i)
\end{equation}where  $\beta = \gamma = 1/|\mathbf{v_t}|$ and  $\delta = 1/|\mathbf{v_t}|(|\mathbf{v_t}|-1)$. Using Eq (5), the original high-dimensional mutual information based standard can be decomposed into a set of low-dimensional MI quantities. We apply this low-dimensional approximation to the selection of lagged variables. 

\noindent{\it {2.2.2. Mixed Search Strategy }}   

An applicable search strategy is important for building the mixed embedding vector. Because the greedy search strategy has high computational efficiency and good practicability, it has become the preferred strategy for embedding. However, the greedy strategy uses a sequential forward approach to select lagged variables, which rely heavily on the  initial embedded vector. That is to say, the initial embedded vector is not accurate and the subsequent selection will get worse.

In order to solve the above problem, we propose a mixed strategy to avoid inaccuracies in the initial embedding. The mixed strategy consists of two strategies: the traversal strategy and the greedy strategy. The application of the strategy is determined by defining a strategy adjustment factor $m$. Assuming that a number of iterations is $k$, the traversal strategy is applied when $1<k \leq m$. For example, when using the traversal strategy, it is necessary to calculate the possible combinations of all lagged variables before determining the mixed embedding vector of the current step. That is to say, we need to calculate $C_{K*L}^k$ combinations in total, and then select the combination of largest conditional mutual information as the mixed embedding vector of the current step. The greedy strategy is applied when $k>m$. This strategy is the same as the one used by the PMIME method.

\noindent{\it {2.2.3. LM-PMIME Method }}   

We put forward the LM-PMIME method for estimating the directional coupling in multivariate time series according to the low-dimensional approximation of conditional mutual information and the mixed search strategy. In the LM-PMIME method, the mixed strategy determines the way to select lagged variables. But whether the variable will be embedded depends on the low dimensional approximation of conditional mutual information. Fig.1 is an illustration of the flow of the LM-PMIME method.

The detailed LM-PMIME method is as follows:
\begin{enumerate}
\item Initialize an empty embedding vector $\mathbf{v{_t^0}} = \emptyset$.
\item At the first iteration $k$ = 1, the embedding vector $w{_t^1}\in W_t$ is selected most related to $\mathbf{y_t}$ :\begin{equation}
w{_t^1}=\mathop{\arg\max}_{w\in W_t}I(\mathbf{y_t} ; w)
\end{equation}Then we have $\mathbf{v{_t^1}} = [w{_t^1}]$. 
\item At the iteration $1<k \leq m $, $w{_t^k}$ will be tested by a standard through computing the maximum value of the low dimensional approximation of CMI. \begin{equation} \begin{split} 
w{_t^k}\!=\!\mathop{\arg\max}_{w\in W_t}I(w;\mathbf{y_t})\!-\!\beta \sum_{w_i \in  \mathbf{v{_t^{k-1}}} } I(w;w_i)\!+\!\gamma\sum_{w_i \in  \mathbf{v{_t^{k-1}}} } I(w;w_i|\mathbf{y_t})\!-\!\delta \sum _{w_i^{ } \in \mathbf{v{_t^{k-1}}} }\sum_{ w_j^{ } \in \mathbf{v{_t^{k-1}}};i \not=j} I(w;w_j|w_i)
\end{split} \end{equation}where  $\beta = \gamma = 1/|\mathbf{v_t}|$ and  $\delta = 1/|\mathbf{v_t}|(|\mathbf{v_t}|-1)$. The traversal strategy is applied to select $\mathbf{v{_t^k}}$, i.e., at the iteration $k=4$ and $m=5$, $\mathbf{v{_t^4}}$ needs to be selected. First, clear the already embedded vector $\mathbf{v{_t^3}}$ and calculate $C_{K*L}^4$ combinations in total. Then select the combination of largest conditional mutual information as $\mathbf{v{_t^4}}$ of the current step. Finally, $k=k+1$.
\item At the iteration $k>m$, greedy strategy is used. Each $w{_t^k}$ will be embedded in the already embedded vector $\mathbf{v{_t^{k-1}}}$ until the process stops. The standard of low dimensional approximation is still used before stopping.
\item The termination criterion is quantified as：\begin{equation}
I(\mathbf{y_t} ; \mathbf{v{_t^{k-1}}})/I(\mathbf{y_t} ; \mathbf{v{_t^k}}) >  A
\end{equation}where the threshold $A < 1$ and threshold $A$ near 1, e.g. $A = 0.95$, allows the inclusion of a new component in the mixed embedding vector even if the augmented vector explains very little of the information on $\mathbf{y}_{\mathbf{t}}$ that was not explained at the previous step. The general value of $A$ is 0.95 or 0.97 in \cite{ref1-9,ref1-10}. That is, the additional information of $w{_t^k}$ selected at the iteration $k$ is not large enough. The embedding process will stop and we have the mixed embedding vector $\mathbf{v_t}=\mathbf{v{_t^{k-1}}}$. In addition, any combination of the lagged variables $X,Y,Z_1,\ldots,Z_{K-2}$ may be included in $\mathbf{v_t}$. 
\item To quantify the causality strength of $X$ on $Y$ conditioned by the other variables in $Z$, the index is defined as :\begin{equation}
R{_{X\rightarrow Y|Z}} = \frac{I(\mathbf{y_t};\mathbf{v{_t^x}}|\mathbf{v{_t^y}},\mathbf{v{_t^z}})}{I(\mathbf{y_t} ; \mathbf{v{_t}})}
\end{equation}where $\mathbf{v{_t^x}}$ represents the component of $X$ in $\mathbf{v{_t}}$. And it is the same with $\mathbf{v{_t^y}}$ and $\mathbf{v{_t^z}}$. The causality strength of $X$ to $Y$ depends on the components of $X$ in $\mathbf{v}_{\mathbf{t}}^{\mathbf{x}}$. The presence of components of $X$ in the mixed embedding vector indicates that $X$ has some effect on the evolution of $Y$ and then the derived information measure causality strength  $R_{X \rightarrow Y | Z}$ is positive, whereas the absence indicates no effect and then causality strength $R_{X \rightarrow Y | Z}$ is exactly zero. In addition, the $R_{X \rightarrow Y | Z}$ is considered significant if it is positive in the PMIME method and proposed method. 
\end{enumerate}
\begin{center}
\includegraphics[width=37em, height=26em]{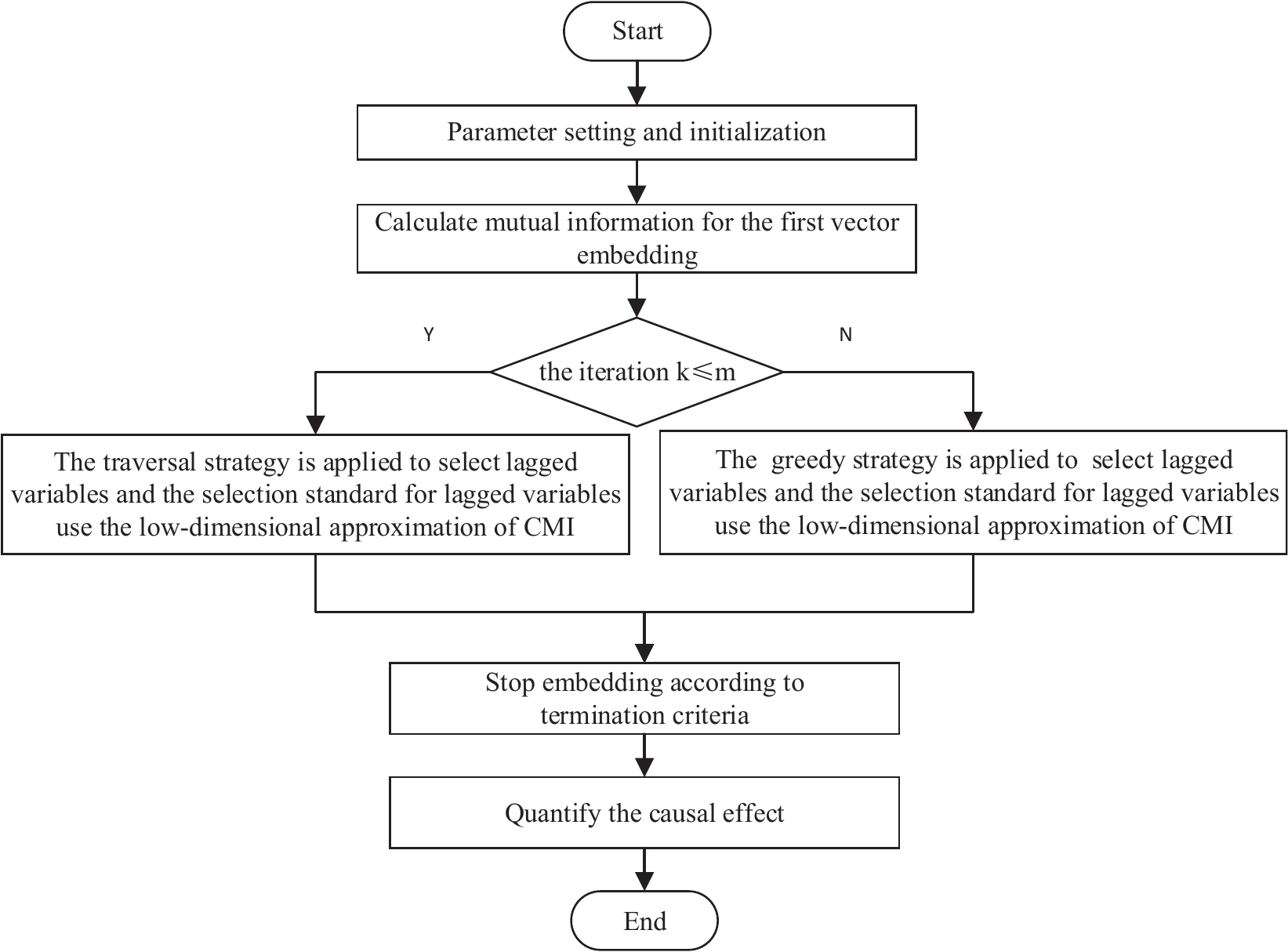}
\end{center}
\begin{center}
\noindent {\small \textbf{Figure 1} The Flow chart of the LM-PMIME method.}
\end{center}

\bigskip\medskip
\noindent {\bf 3. Simulation study}
\smallskip

In this section, we perform a series of causal analysis simulation experiments using linear stochastic, nonlinear stochastic, and chaotic systems. The experiments compare the differences between the proposed LM-PMIME method and the traditional PMIME method for time series with different lengths or coupling strengths. The experiments also add a comparison method M-PMIME, which improve the search strategy without using low-dimensional approximation.

We calculate all methods on 100 realizations from each system to assess statistically the sensitivity and specificity of the methods. The connections between variables are classified as coupled directions and uncoupled directions to compute the confusion matrix: $TP$ (ture positives), $FP$ (false positives), $TN$(true negatives), and $FN$ (false negatives), where sensitivity = $TP/(TP + FN)$, specificity = $TN/(TN + FP)$, and F1 score = $2TP/(2TP + FP+ FN)$.

The accuracy of the estimated mutual information is vital for embedding vector selection \cite{ref2-23}. The two most common methods for estimating mutual information are the histogram and kernel methods. The former one is time efficient but not highly accurate \cite{ref3-26}. The latter one has higher accuracy but comes with huge computational pressure \cite{ref3-27}. We applied the k-nearest neighbors (k-NNs) method to estimate mutual information, because the k-NNs estimator is suitable for high-dimensional data \cite{ref3-28}.

In the following results, the performance of LM-PMIME and a comparison to M-PMIME and PMIME are presented for multivariate time series with different lengths or coupling strengths.

\noindent{\it {3.1. Linear multivariate stochastic process }}

The first system is a linear vector autoregressive (VAR) process which is composed of order 4 in 5 time series (model 1 in \cite{ref3-29}). \begin{equation}
\left\{                        
\begin{aligned}
x_{1,t}  = 0.4x_{1,t-1} - 0.5x_{1,t-2} + 0.4x_{5,t-1} + e_{1,t}\\
x_{2,t}  = 0.4x_{2,t-1} - 0.3x_{1,t-4} + 0.4x_{5,t-2} + e_{2,t}\\
x_{3,t}  = 0.5x_{3,t-1} - 0.7x_{3,t-2} - 0.3x_{5,t-3} + e_{3,t}\\
x_{4,t}  = 0.8x_{4,t-3} + 0.4x_{1,t-2} + 0.3x_{2,t-2} + e_{4,t}\\
x_{5,t}  = 0.7x_{5,t-1} - 0.5x_{5,t-2} - 0.4x_{4,t-1} + e_{5,t}\\
\end{aligned}
\right.
\end{equation}where $e_{i,t}$ , $i = 1,\cdots\ , 5$, are Gaussian noise with zero mean and unit covariance matrix. $X_1\rightarrow X_2$, $X_1\rightarrow X_4$, $X_2\rightarrow X_4$, $X_4\rightarrow X_5$, $X_5\rightarrow X_1$, $X_5\rightarrow X_2$, and $X_5\rightarrow X_3$  are the true causal connections in this process.
\begin{center}
\includegraphics[width=46em, height=13em]{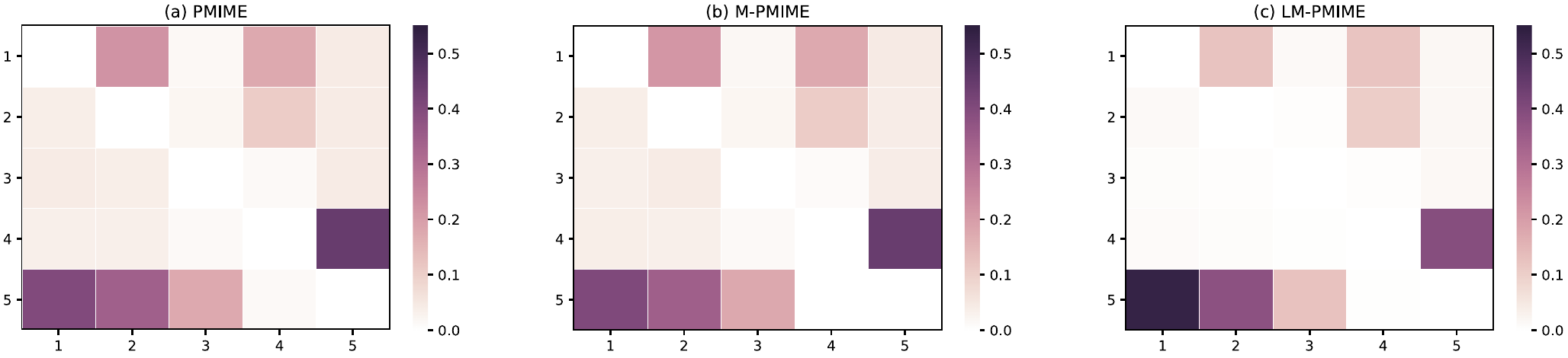}
\end{center}
\noindent {\small \textbf{Figure 2} Matrix representation of causality for the linear VAR process. Retrieved by traditional PMIME method (a), M-PMIME method (b), and LM-PMIME method (c) with k-NNs estimator. The length of the time series is 512. $m=2$ is used for the M-PMIME method and the LM-PMIME method. The remaining parameters of the three methods are the same ($L=6, A=0.97$). Color maps for the mean values of coupling measurements are obtained from 100 realizations of the linear VAR process. The direction of causal influence is from row to column in the matrix. The true causal connections in this linear VAR process are at the matrix elements (1, 2), (1, 4), (2, 4), (4, 5), (5, 1), (5, 2) and (5, 3).}
\\

We use $A=0.97$ and $L = 6$, which matches the larger lag for the three methods in the process. In addition, the LM-PMIME method and the M-PMIME method use the parameter $m=2$. The results from linear VAR process with the time series length of 512 are shown in Fig 2. The direction of causal influence is from row to column in the matrix representation, e. g. the causal connection  $X_1\rightarrow X_2$ is represented as (1, 2) in the matrix representation. Hence, true causal connections in this process are at the matrix elements (1, 2), (1, 4), (2, 4), (4, 5), (5, 1), (5, 2) and (5, 3). The mean values of coupling measured by the three methods are positive and high on these matrix elements. It is proved that the three methods have good sensitivity to true couplings. However, Fig 2 shows that there are lots of false positives in the traditional method using high-dimensional CMI. In contrast, the LM-PMIME method has better performance than the other two methods, because the method reduces false positives. The sensitivity, specificity, and F1 score are obtained from 100 realizations of linear VAR process with varying length of time series. The values of the specific indexs are listed in Table I. The F1 score of LM-PMIME method has better results on linear VAR process with different time series lengths. Furthermore, the F1 score calculated by the LM-PMIME method increases as the length of the time series increases. These better results are likely due to the great improvement of specificity by the proposed method. At the same time, the F1 score reflects that PMIME method and M-PMIME method have achieved similar results in the linear VAR process. It shows that the mixed strategy does not work in this process. However, the following experiments show that the mixed strategy works well on the chaotic system.

\noindent {\small \textbf{TABLE I} Sensitivity, specificity, and F1 score are obtained from 100 realizations of linear VAR process with varying length of time series for the three different methods. $A=0.97$ and $L = 6$ are the parameters common to the three methods. In addition, the LM-PMIME method and the M-PMIME method use the parameter $m=2$.}
\begin{center}
\begin{tabular}{cccccc}
  \hline
  \hline
    $~$               & Sensitivity    &   Specificity & F1 score
    \\
   \hline
   $~$            &  $~$     &   $n = 256$     & $~$  
  \\
   PMIME          & $0.988$   &   $0.492$     & $0.600$
  \\
   M-PMIME        & $0.989$   &   $0.481$     & $0.596$
  \\
   LM-PMIME       & $0.797$   &   $0.741$     & $0.647$
   \\
   $~$            &  $~$     &   $n = 512$     & $~$  
  \\
   PMIME          & $1.000$   &   $0.567$     & $0.643$
  \\
   M-PMIME        & $0.994$   &   $0.727$     & $0.645$
  \\
   LM-PMIME       & $0.855$   &   $0.763$     & $0.693$
    \\
   $~$            &  $~$     &   $n = 1024$     & $~$  
  \\
   PMIME          & $1.000$   &   $0.697$     & $0.719$
  \\
   M-PMIME        & $0.940$   &   $0.729$     & $0.713$
  \\
   LM-PMIME       & $0.877$   &   $0.807$     & $0.739$
  \\
  \hline
  \hline
\end{tabular}
\end{center}

\noindent{\it {3.2. Nonlinear multivariate stochastic process}}

The nonlinear VAR process is of order 1 in three variables $\mbox{NLVAR}_3(1)$ (model 7 in \cite{ref3-31}).\begin{equation}
\left\{
\begin{array}{l}
x_{1,t} = 3.4x_{1,t-1}(1-x_{1,t-1}^2)e^{-x_{1,t-1}^2} + 0.4e_{1,t}\\
x_{2,t} = 3.4x_{2,t-1}(1-x_{2,t-1}^2)e^{-x_{2,t-1}^2} + 0.5x_{1,t-1}x_{2,t-1} + 0.4e_{2,t}\\
x_{3,t} = 3.4x_{3,t-1}(1-x_{3,t-1}^2)e^{-x_{3,t-1}^2} + 0.3x_{2,t-1} + 0.5x_{1,t-1}^2 + 0.4e_{3,t}\\
\end{array}
\right.
\end{equation}The true causal connections in NLVAR3 (1) are $X_1\rightarrow X_2$, $X_1\rightarrow X_3$, $X_2\rightarrow X_3$. The results obtained from 100 realizations of the nonlinear VAR process are shown in Fig 3 for $n = 512$, $A=0.97$, $L=6$. The strategy adjustment factor $m=3$ determines the application of the strategies for LM-PMIME method and M-PMIME method. The true causal connections in $\mbox{NLVAR}_3(1)$ are represented at the matrix elements (1,2), (1,3), and (2,3). For the three methods, the mean values of coupling measurements on these matrix elements are positive and high. It turns out that all methods have good sensitivity to true couplings. But there are many false positives in the traditional methods using high-dimensional CMI. Hence, the LM-PMIME method significantly outperforms the others. The sensitivity, specificity, and F1 score are obtained from $\mbox{NLVAR}_3(1)$ by gradually increasing the time series length from 256 to 1024. The values of the specific indexs are listed in Table II. The F1 score of LM-PMIME method has better results on $\mbox{NLVAR}_3(1)$ with different time series lengths. In addition, the F1 score will increase as the length of the time series increases. The low-dimensional approximation of CMI can greatly improve specificity, although mixed strategy does not work in $\mbox{NLVAR}_3(1)$.
\begin{center}
\includegraphics[width=46em, height=13em]{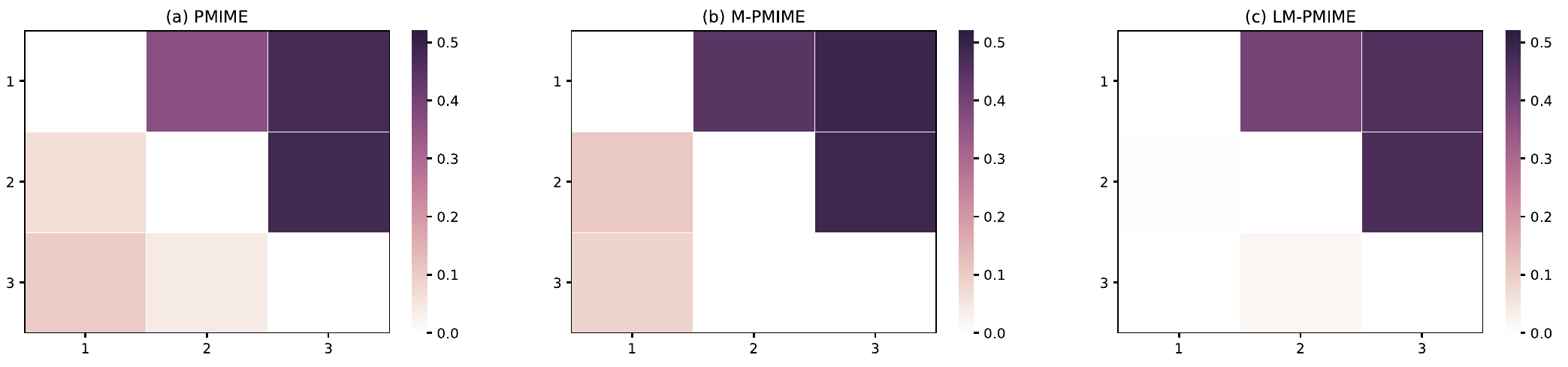}
\end{center}
\noindent {\small \textbf{Figure 3} Matrix representation of causality for $\mbox{NLVAR}_3(1)$. Retrieved by traditional PMIME method (a), M-PMIME method (b), and LM-PMIME method (c) with k-NNs estimator. The length of the time series is 512. $m=3$ is used for the M-PMIME method and the LM-PMIME method. The remaining parameters of the three methods are the same ($L=6$, $A=0.97$).  Color maps for the mean values of coupling measurements are obtained from 100 realizations of $\mbox{NLVAR}_3(1)$. The direction of causal influence is from row to column in the matrix. The true causal connections in $\mbox{NLVAR}_3(1)$ are at the matrix elements (1,2), (1,3), (2,3).}
\\
\noindent {\small \textbf{TABLE II}  Sensitivity, specificity, and F1 score are obtained from 100 realizations of $\mbox{NLVAR}_3(1)$ with varying length of time series for the three different methods. $A=0.97$ and $L = 6$ are the parameters common to the three methods. In addition, the LM-PMIME method and the M-PMIME method use the parameter $m=3$.}
\begin{center}
\begin{tabular}{cccccccc}
  \hline
  \hline
    $~$               & Sensitivity    &   Specificity & F1 score
    \\
   \hline
   $~$            &  $~$     &   $n = 256$     & $~$  
  \\
   PMIME          & $0.973$   &   $0.650$     & $0.737$
  \\
   M-PMIME        & $0.976$   &   $0.615$     & $0.712$
  \\
   LM-PMIME       & $0.860$   &   $0.844$     & $0.792$
   \\
   $~$            &  $~$     &   $n = 512$     & $~$  
  \\
   PMIME          & $1.000$   &   $0.681$     & $0.758$
  \\
   M-PMIME        & $1.000$   &   $0.662$     & $0.748$
  \\
   LM-PMIME       & $0.950$   &   $0.887$     & $0.873$
    \\
   $~$            &  $~$     &   $n = 1024$     & $~$  
  \\
   PMIME          & $1.000$   &   $0.860$     & $0.877$
  \\
   M-PMIME        & $1.000$   &   $0.790$     & $0.827$
  \\
   LM-PMIME       & $0.989$   &   $0.892$     & $0.896$
  \\
  \hline
  \hline
\end{tabular}
\end{center}

\noindent{\it {3.3. Coupled Henon maps}}

The system of $K$ coupled chaotic Henon maps defined as\begin{equation}
\left\{
\begin{array}{l}
x_{1,t} = 1.4-x_{1,t-1}^2+0.3x_{1,t-2}\\
x_{i,t} = 1.4-(Cx_{i-1,t-1}+(1-C)x_{i,t-1})^2+0.3x_{i,t-2}  \qquad  for\; i = 2,\cdots\ , K\\
\end{array}
\right.
\end{equation}$X_{i-1}\rightarrow X_{i}$, where $i = 2,\cdots\ , K$, are the true causal connections in the $K$ coupled chaotic Henon maps. The results from 100 realizations of the coupled Henon maps with the coupling strength $C=0.1$ are shown in Fig 4 for $n = 1024$, $K=6$, $A=0.95$, $L=5$, $m=2$. In addition to this, the results of only changing the coupling strength $C=0.3$ are shown in Fig 5. The true causal connections in the coupled Henon maps are at the matrix elements ($i-1, i$), where $i = 2,\cdots\ , 6$. There is almost no false positive for all methods. However, Fig 4 and Fig 5 illustrate that the proposed methods have better performance than the traditional method when there are true causal connections. All methods will detect stronger causal connections as the coupling strength $C$ of the system increases. The sensitivity, specificity, and F1 score are obtained from coupled Henon maps with the variables $K$ from 3 to 9. The values of the specific indexs are listed in Table III and Table IV. The results show that the F1 score of the LM-PMIME method is higher than the others when the coupling strength is low. Although the F1 score may be affected by the number of variables $K$ in the simulation experiments, the F1 score for the LM-PMIME method is above 0.9. The LM-PMIME method and the M-PMIME method greatly improve the specificity, especially the former method. It is proved that both low-dimensional approximation of CMI and the mixed strategies play an important role in coupled Henon maps when the coupling strength is low.
\begin{center}
\includegraphics[width=46em, height=13em]{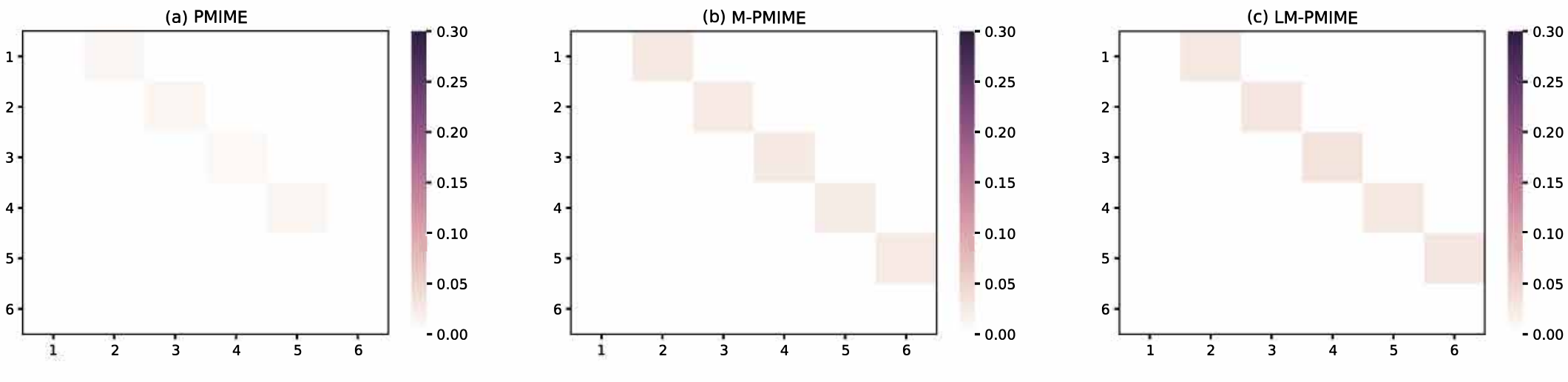}
\end{center}
\noindent {\small \textbf{Figure 4} Matrix representation of causality for $K = 6$ variables of the coupled Henon maps ($C=0.1$). Retrieved by traditional PMIME method (a), M-PMIME method (b), and LM-PMIME method (c) with k-NNs estimator. The length of the time series is 1024. $m=2$ is used for the M-PMIME method and the LM-PMIME method. The remaining parameters of the three methods are the same ($L=5$, $A=0.95$).  Color maps for the mean values of coupling measurements are obtained from 100 realizations of the coupled Henon maps. The direction of causal influence is from row to column in the matrix. The true causal connections in the coupled Henon maps are at the matrix elements ($i-1, i$), where $i = 2, \cdots\ , 6$.}
\\
\begin{center}
\includegraphics[width=46em, height=13em]{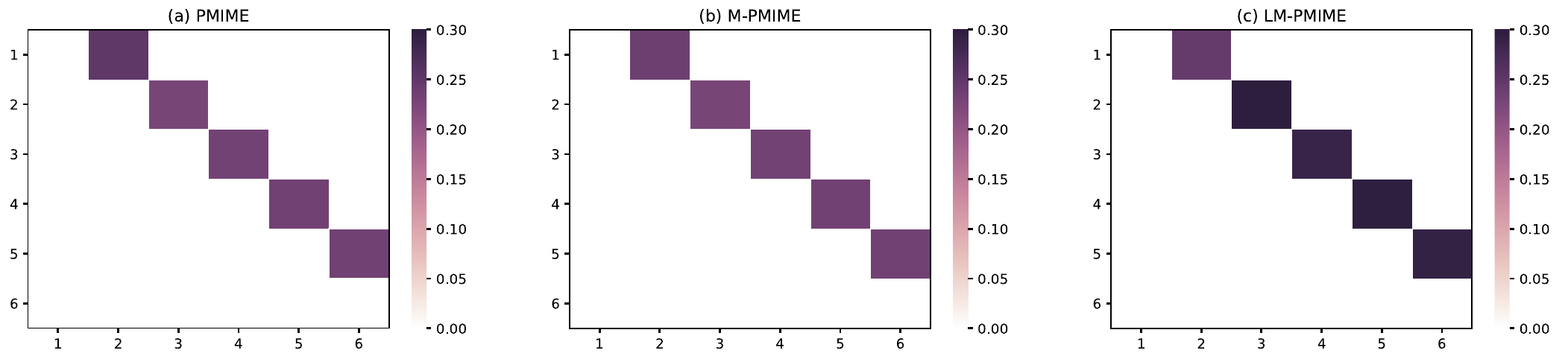}
\end{center}
\noindent {\small \textbf{Figure 5} Matrix representation of causality for $K = 6$ variables of the coupled Henon maps ($C=0.3$). Retrieved by traditional PMIME method (a), M-PMIME method (b), and LM-PMIME method (c) with k-NNs estimator. The length of the time series is 1024. $m=2$ is used for the M-PMIME method and the LM-PMIME method. The remaining parameters of the three methods are the same ($L=5$, $A=0.95$).  Color maps for the mean values of coupling measurements are obtained from 100 realizations of the coupled Henon maps. The direction of causal influence is from row to column in the matrix. The true causal connections in the coupled Henon maps are at the matrix elements ($i-1, i$), where $i = 2, \cdots\ , 6$.}
\\

\noindent {\small \textbf{TABLE III} Sensitivity, specificity, and F1 score are obtained from 100 realizations of $K$ variables of the coupled Henon maps ($C=0.1$) for the three different methods. $A=0.95$ and $L = 5$ are the parameters common to the three methods. In addition, the LM-PMIME method and the M-PMIME method use the parameter $m=2$.}
\begin{center}
\begin{tabular}{ccccccccc}
  \hline
  \hline
    $~$               & Sensitivity    &   Specificity & F1 score
    \\
   \hline
   $~$            &  $~$     &   $K = 3$     & $~$  
  \\
   PMIME          & $0.175$   &   $1.000$     & $0.297$
  \\
   M-PMIME        & $0.715$   &   $1.000$     & $0.834$
  \\
   LM-PMIME       & $0.945$   &   $1.000$     & $0.972$
   \\
   $~$            &  $~$     &   $K = 6$     & $~$  
  \\
   PMIME          & $0.217$   &   $1.000$     & $0.357$
  \\
   M-PMIME        & $0.674$   &   $1.000$     & $0.806$
  \\
   LM-PMIME       & $0.926$   &   $0.998$     & $0.950$
    \\
   $~$            &  $~$     &   $K = 9$     & $~$  
  \\
   PMIME          & $0.204$   &   $1.000$      & $0.338$
  \\
   M-PMIME        & $0.700$   &   $1.000$       & $0.824$
  \\
   LM-PMIME       & $0.895$   &   $0.998$     & $0.904$
  \\
  \hline
  \hline
\end{tabular}
\end{center}

\noindent {\small \textbf{TABLE IV}Sensitivity, specificity, and F1 score are obtained from 100 realizations of $K$ variables of the coupled Henon maps ($C=0.3$) for the three different methods. $A=0.95$ and $L = 5$ are the parameters common to the three methods. In addition, the LM-PMIME method and the M-PMIME method use the parameter $m=2$.}
\begin{center}
\begin{tabular}{cccccccccc}
  \hline
  \hline
    $~$               & Sensitivity    &   Specificity & F1 score
    \\
   \hline
   $~$            &  $~$     &   $K = 3$     & $~$  
  \\
   PMIME          & $1.000$   &   $1.000$     & $1.000$
  \\
   M-PMIME        & $1.000$   &   $1.000$     & $1.000$
  \\
   LM-PMIME       & $1.000$  &   $1.000$     & $1.000$
   \\
   $~$            &  $~$     &   $K = 6$     & $~$  
  \\
   PMIME          & $1.000$  &   $1.000$     & $1.000$
  \\
   M-PMIME        & $1.000$   &   $1.000$     & $1.000$
  \\
   LM-PMIME       & $1.000$   &   $1.000$    & $1.000$
    \\
   $~$            &  $~$     &   $K = 9$     & $~$  
  \\
   PMIME          & $1.000$   &   $1.000$      & $1.000$
  \\
   M-PMIME        & $1.000$   &   $1.000$       & $1.000$
  \\
   LM-PMIME       & $1.000$   &   $1.000$     & $1.000$
  \\
  \hline
  \hline
\end{tabular}
\end{center}

\noindent{\it {3.4. Coupled Lorenz system}}

Next we study a chaotic system of three coupled identical Lorenz oscillators defined as\begin{equation}
\left\{
\begin{array}{l}
\dot{x_1} = -10x_1 + 10y_1,  \quad \dot{x_i} = -10x_i + 10y_i + C (x_{i-1} - x_i),\\
\dot{y_1} = -x_1z_1 + 28x_1 - y_1,  \quad  \dot{y_i} = -x_i z_i + 28x_i - y_i,\\
\dot{z_1} = x_1y_1 - \frac{8}{3}z_1,  \quad \dot{z_i} = x_i y_i - \frac{8}{3}z_i,\\
\end{array}
\right.
\end{equation}where $i=2,3$. The differential equations by the explicit Runge-Kutta (4,5) method are solved in MATLAB. In addition, the time series are generated at a sampling time of 0.05 time units. The true causal connections in the three coupled Lorenz oscillators are $X_{i-1}\rightarrow X_{i}$, where $i=2,3$.

The results from 100 realizations of the three coupled Lorenz oscillators with the coupling strength $C=3$ are shown in Fig 6, for $n = 512$, $A=0.95$, $L=5$, $m=3$. In addition, The sensitivity, specificity, and F1 score are listed in Table V. The values of the specific indexs are obtained from the three coupled Lorenz oscillators with varying length of time series from 256 to 1024 and the remaining parameters are the same. The F1 scores of the proposed methods are much higher than the traditional PMIME method. The M-PMIME method performs best when the time series is short. That is to say, the mixed strategy plays a role in improving the F1 score. However, the F1 score of the LM-PMIME method is the highest as the length of the time series increases.
\begin{center}
\includegraphics[width=46em, height=13em]{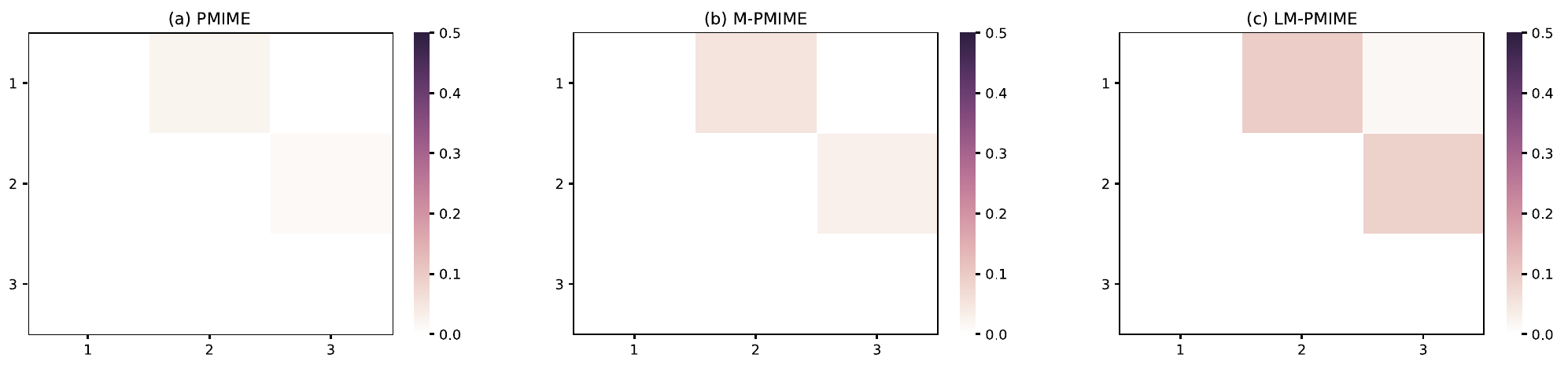}
\end{center}
\noindent {\small \textbf{Figure 6} Matrix representation of causality for the three coupled Lorenz oscillators. Retrieved by traditional PMIME method (a), M-PMIME method (b), and LM-PMIME method (c) with k-NNs estimator. The length of the time series is 512 with coupling strength $C=3$. $m=3$ is used for the M-PMIME method and the LM-PMIME method. The remaining parameters of the three methods are the same ($L=5$, $A=0.95$). Color maps for the mean values of coupling measurements are obtained from 100 realizations of the three coupled Lorenz oscillators. The direction of causal influence is from row to column in the matrix. The true causal connections in the three coupled Lorenz oscillators are at the matrix elements ($i-1, i$), where $i=2,3$.}
\\
\noindent {\small \textbf{TABLE V} Sensitivity, specificity, and F1 score are obtained from 100 realizations of the three coupled Lorenz oscillators ($C=3$) with varying length of time series for the three different methods. $A=0.95$ and $L = 5$ are the parameters common to the three methods. In addition, the LM-PMIME method and the M-PMIME method use the parameter $m=3$.}
\begin{center}
\begin{tabular}{cccccccccc}
  \hline
  \hline
    $~$               & Sensitivity    &   Specificity & F1 score
    \\
   \hline
   $~$            &  $~$     &   $n = 256$     & $~$  
  \\
   PMIME          & $0.225$   &   $0.997$     & $0.364$
  \\
   M-PMIME        & $0.660$   &   $0.863$     & $0.617$
  \\
   LM-PMIME       & $0.805$   &   $0.665$     & $0.541$
   \\
   $~$            &  $~$     &   $n = 512$     & $~$  
  \\
   PMIME          & $0.185$   &   $1.000$     & $0.312$
  \\
   M-PMIME        & $0.640$   &   $0.913$     & $0.658$
  \\
   LM-PMIME       & $0.875$   &   $0.743$     & $0.631$
    \\
   $~$            &  $~$     &   $n = 1024$     & $~$  
  \\
   PMIME          & $0.175$   &   $1.000$     & $0.297$
  \\
   M-PMIME        & $0.670$   &   $0.909$     & $0.674$
  \\
   LM-PMIME       & $0.970$   &   $0.756$     & $0.687$
  \\
  \hline
  \hline
\end{tabular}
\end{center}

The sensitivity, specificity, and F1 score are obtained from 100 realizations of the three coupled Lorenz oscillators with coupling strength $C$ from 1 to 5 for the three different methods. The length of the time series is 512 and $A=0.95$, $L=5$, $m=3$. The values of sensitivity, specificity, and F1 score are listed in Table VI. The results show that the LM-PMIME method performs best when the coupling strength $C$ is low, such as $C$ =1. Although the F1 score of the traditional PMIME method increases as the coupling strength $C$ increases, it is still much worse than the proposed methods. Fig 7 is the matrix representation of causality for the three coupled Lorenz oscillators with coupling strength $C=5$. The true causal connections are ($i-1, i$) in the matrix elements, where $i=2,3$. Only for the LM-PMIME method, the mean values of coupling measurements on these matrix elements are positive and high.
\begin{center}
\includegraphics[width=46em, height=13em]{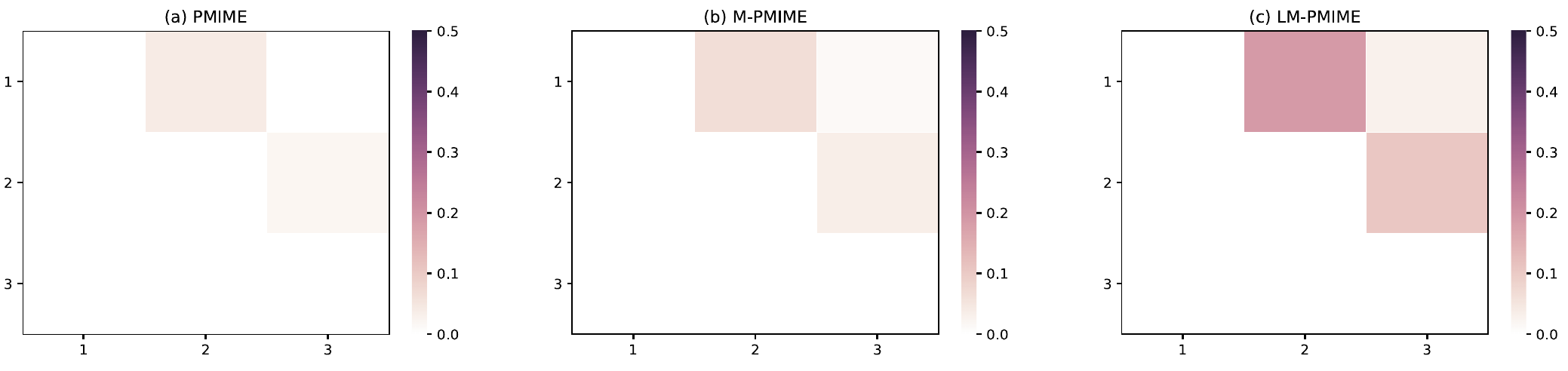}
\end{center}
\noindent {\small \textbf{Figure 7} Matrix representation of causality for the three coupled Lorenz oscillators. Retrieved by traditional PMIME method (a), M-PMIME method (b), and LM-PMIME method (c) with k-NNs estimator. The length of the time series is 512 with coupling strength $C=5$. $m=3$ is used for the M-PMIME method and the LM-PMIME method. The remaining parameters of the three methods are the same ($L=5$, $A=0.95$). Color maps for the mean values of coupling measurements are obtained from 100 realizations of the three coupled Lorenz oscillators. The direction of causal influence is from row to column in the matrix. The true causal connections in the three coupled Lorenz oscillators are at the matrix elements ($i-1, i$), where $i=2,3$.}
\\
\noindent {\small \textbf{TABLE VI} Sensitivity, specificity, and F1 score are obtained from 100 realizations of the three coupled Lorenz oscillators ($n=512$) with coupling strength $C$ from 1 to 5 for the three different methods. $A=0.95$ and $L = 5$ are the parameters common to the three methods. In addition, the LM-PMIME method and the M-PMIME method use the parameter $m=3$.}
\begin{center}
\begin{tabular}{ccccccccccc}
  \hline
  \hline
    $~$               & Sensitivity    &   Specificity & F1 score
    \\
   \hline
   $~$            &  $~$     &   $C = 1$     & $~$  
  \\
   PMIME          & $0.000$   &   $1.000$     & $0.000$
  \\
   M-PMIME        & $0.155$   &   $0.926$     & $0.221$
  \\
   LM-PMIME       & $0.375$   &   $0.830$     & $0.381$
   \\
   $~$            &  $~$     &   $C = 2$     & $~$  
  \\
   PMIME          & $0.075$   &   $1.000$     & $0.141$
  \\
   M-PMIME        & $0.565$   &   $0.893$     & $0.583$
  \\
   LM-PMIME       & $0.825$   &   $0.765$     & $0.623$
    \\
   $~$            &  $~$     &   $C = 3$     & $~$  
  \\
   PMIME          & $0.185$   &   $1.000$     & $0.312$
  \\
   M-PMIME        & $0.640$   &   $0.913$     & $0.658$
  \\
   LM-PMIME       & $0.875$   &   $0.743$     & $0.631$
    \\
     &  $~$     &   $C = 4$     & $~$  
  \\
   PMIME          & $0.260$   &   $1.000$     & $0.413$
  \\
   M-PMIME        & $0.740$   &   $0.892$     & $0.698$
  \\
   LM-PMIME       & $0.920$   &   $0.710$     & $0.627$
  \\
     &  $~$     &   $C =5$     & $~$  
  \\
   PMIME          & $0.320$   &   $0.997$     & $0.481$
  \\
   M-PMIME        & $0.725$   &   $0.873$     & $0.660$
  \\
   LM-PMIME       & $0.960$   &   $0.731$     & $0.661$
  \\
  \hline
  \hline
\end{tabular}
\end{center}

\bigskip\medskip
\noindent {\bf 4. Application}
\smallskip

In this section, we show the applicability of the proposed LM-PMIME method to actual electrocorticographic (ECoG) data. That is to say, the causal analysis method is adopted to explore key contacts of the human subject with intractable epilepsy and assist doctors in the diagnosis and treatment of the disease. A public dataset from the human subject (a 39-year-old woman with medically refractory complex partial seizures) is used. The dataset contains 8 seizure epochs and 8 pre-seizure epochs. Each epoch contains 76 time series obtained from the 8-by-8 electrode grid and two depth electrodes with six contacts each. In addition, the duration of each epoch is 10s and the length of each time series is 4000 (More details about the data are given in \cite{ref4-33}).

We use PMIME method and LM-PMIME method to analyze the seizure data and the pre-seizure data. The data is recorded at a fixed sampling rate of 400 Hz, which is downsampled to 100 Hz. To assess the causal matrices of different physiological states estimated by each method, we compute the average causal strengths (the mean values of the coupling measurements over all epochs in the same physiological state) as shown in Fig. 8. The brighter the colors are, the more signifincant causal connections are. As a result, it is obvious from the causal matrics of LM-PMIME method that contact 73 has more impact on the other contacts, highlighting that it is the key contact in the pre-seizure data [see Fig. 8(b)]. However, the traditional PMIME method has led to a lot of false positives [see Fig. 8(a)]. Fig. 9 illustrates the difference of total numbers of significant connections between the seizure state and the pre-seizure state. The proposed method highlights the key contact 50  [see Fig. 9(b)] and these discovered key contacts are consistent with many researchers \cite{ref4-33,ref4-34,ref4-35}.

\begin{center}
\includegraphics[width=40em, height=30em]{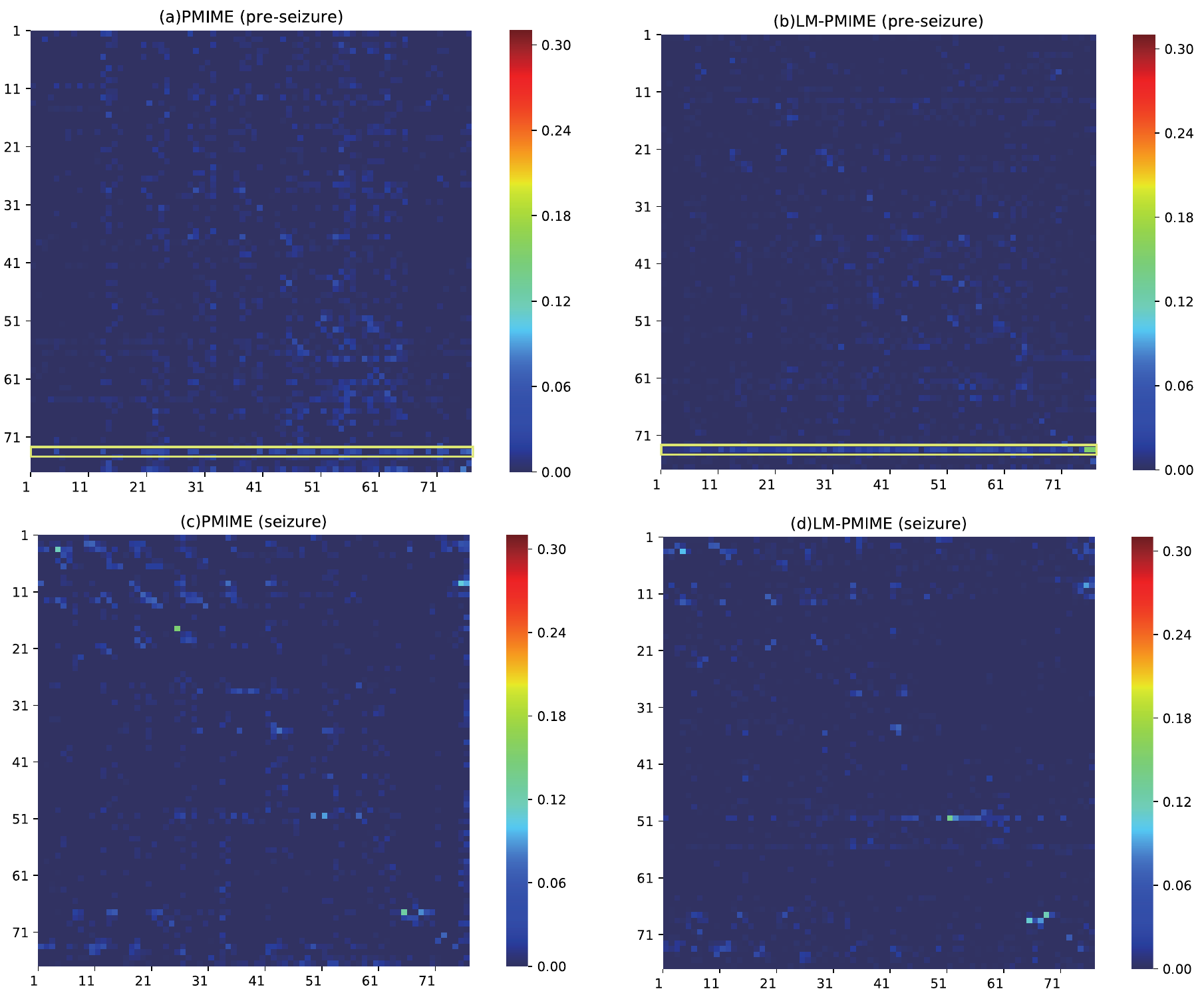}
\end{center}
\noindent {\small \textbf{Figure 8} Results for multivariate electrocorticographic (ECoG) data. Matrices of causalities reflect the pre-seizure state (top) and the seizure state (down) estimated by the PMIME method and the LM-PMIME method. The causal strengths are averaged (the mean values of the coupling measurements over all epochs in the same physiological state). Contacts 1 to 64 belong to an 8-by-8 electrode grid, and contacts 65 to 76 belong to two depth electrodes. The direction of causal influence is from row to column in the matrices. The brighter colors indicate more significant values. The key contact is marked by a rectangular box. The parameter $A = 0.95$ and $m=2$ are set for the different methods.}

\begin{center}
\includegraphics[width=40em, height=15em]{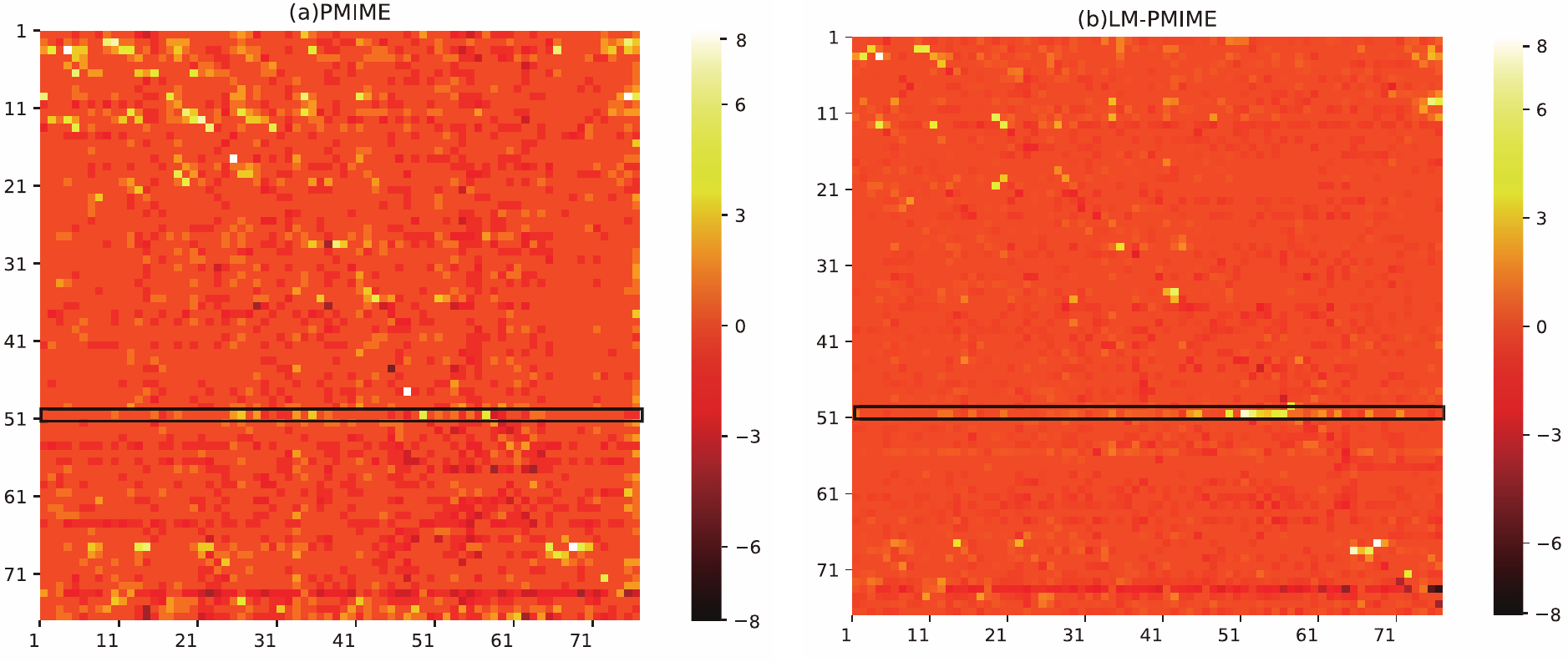}
\end{center}
\noindent {\small \textbf{Figure 9} Results for multivariate electrocorticographic (ECoG) data. Matrices reflect the difference of total numbers of significant connections between the seizure state and the pre-seizure state (seizure minus pre-seizure). The numbers are respectively summed from 8 seizure epochs and 8 pre-seizure epochs. Contacts 1 to 64 belong to an 8-by-8 electrode grid, and contacts 65 to 76 belong to two depth electrodes. The direction of causal influence is from row to column in the matrices. The brighter colors indicate more significant values. The key contact is marked by a rectangular box. The parameter $A = 0.95$ and $m=2$ are set for the different methods.}
 
We observe that LM-PMIME method gives an obvious causal driver located at the contact 73 from the second depth electrode strip in the pre-seizure data. Therefore, the contact 73 may be associated with seizures. Although not yet clinically observable, it has been suggested that the second depth electrode primarily affect cortical activity in \cite{ref4-34,ref4-35}. In addition, the proposed method successfully identifies a key contact from the data: contact 50, which exhibits the most significant change in the betweenness centrality. The contact is considered the primary target of therapeutic intervention in [38], because contacts with statistically significant increases in betweenness centrality may lead to seizures. In contrast, traditional PMIME method leads to a large number of false positives, so key contacts cannot be highlighted.

\bigskip\medskip

\noindent{\bf 5. Discussion and conclusion}

\smallskip

In this paper, we have put forward a new non-uniform embedding method named LM-PMIME for multivariate time series. We present effective modifications for the well-known non-uniform embedding method: PMIME, which quantifies causality by means of information theoretic measures. The advantage of the non-uniform embedding compared with uniform embedding is that it can reduce the dimension of the state space by selecting the relevant variables which contribute the most to explain the target variable. Therefore, it has been proved that the non-uniform embedding process is more flexible for state space reconstruction \cite{ref1-8,ref1-9,ref4-36}. However, there are still some obstacles to estimating high-dimensional CMI and forming optimal mixed embedding vector in the traditional non-uniform embedding methods. The proposed LM-PMIME method overcomes the above shortcomings of traditional methods. The effectiveness and applicability of the LM-PMIME method are demonstrated by a large number of experiments. Furthermore, the LM-PMIME method works well for multivariate time series with weak coupling strengths, especially for chaotic systems. The usefulness of the LM-PMIME method for multivariate time series is illustrated by the analysis of actual ECoG data.

The major contribution of the proposed LM-PMIME method, which is based on the low-dimensional approximation of conditional mutual information and the mixed search strategy, is that improves the traditional non-uniform embedding methods. The curse of dimensionality is avoided by replacing the original estimate with a low-dimensional approximation of conditional mutual information. In addition, a mixed strategy instead of the greedy strategy is used as an embedded strategy to solve the problem of initial embedding inaccuracy. Hence, the mixed embedding vector becomes more parsimonious by maximizing the correlation with the target variable and minimizing the redundancy between the selected variables. In order to form the optimal mixed embedded vector, there are also other propositions. For example, in \cite{ref1-16} a preselection scheme for subsets of causal predictors is used to search an optimal subset and detect the synergetic variables. In addition, many researchers adopt the OCE algorithm \cite{ref-add-1} or the PCMCI \cite{ref-add-2} algorithm to estimate the causal graphs. Different from these preselection methods, the LM-PMIME method relies on both the low-dimensional approximation and the mixed search strategy to improve the conditions. In all simulation systems, the LM-PMIME method performs better than the traditional methods according to the F1 score. Because of the complexity of chaotic systems, true causality is often difficult to detect. However, the LM-PMIME method significantly improves the sensitivity in chaotic systems. In the remaining simulation systems, the LM-PMIME method reduces false positives and increases the specificity. The experiments also adopt the comparison method M-PMIME, which improves the search strategy without using low-dimensional approximation. By the M-PMIME method, it can be found that the mixed search strategy works well in chaotic systems, especially the systems with low coupling strengths. In addition, the low-dimensional approximation of conditional mutual information plays an important role in linear and nonlinear systems. Therefore, we combine both the low-dimensional approximation of conditional mutual information and the mixed search strategy to form a new non-uniform embedding method LM-PMIME for multivariate time series.

In this study, the proposed LM-PMIME method, a causal analysis method, has great potential to be adopted in other applications, e.g. prediction of dynamic systems. We will further study the non-uniform embedding method and extend its applications.

\bigskip\medskip
\begin{center}
\noindent{\bf Acknowledgement}
\end{center}
\smallskip

Financial supports by National Natural Science Foundation of China (61603029), and the Fundamental Research Funds for the Central Universities (2018RC002) are gratefully acknowledge.

\end{document}